# A Review of Network Traffic Analysis and Prediction Techniques

**Manish R. Joshi**[1]
*School of Computer Sciences, North Maharashtra University,*
*Jalgaon (M.S ) India*
*joshmanish@gmail.com*
**Theyazn Hassn Hadi**[2]
*School of Computer Sciences, North Maharashtra University,*
*Jalgaon (M.S ) India*
*thhadi@nmu.ac.in*

**Abstract**

Analysis and prediction of network traffic has applications in wide comprehensive set of areas and has newly attracted significant number of studies. Different kinds of experiments are conducted and summarized to identify various problems in existing computer network applications. Network traffic analysis and prediction is a proactive approach to ensure secure, reliable and qualitative network communication. Various techniques are proposed and experimented for analyzing network traffic including neural network based techniques to data mining techniques. Similarly, various Linear and non-linear models are proposed for network traffic prediction. Several interesting combinations of network analysis and prediction techniques are implemented to attain efficient and effective results.

This paper presents a survey on various such network analysis and traffic prediction techniques. The uniqueness and rules of previous studies are investigated. Moreover, various accomplished areas of analysis and prediction of network traffic have been summed.

## 1. Introduction

Considering the fact that e-commerce, banking and business related highly confidential and valuable information communicated within the network, it is needless to mention the importance of network traffic analysis to attain proper information security. Network traffic analysis and prediction resembles a proactive approach rather than reactive, where network is monitored to ensure that security breaches do not occur within network. The network traffic analysis is a significant stage for developing successful preventive congestion control schemes and to find out normal and malicious packets. These schemes target to avoid network congestion by distributing the network resources with respect to the forecasted traffic.

The predictability of network traffic is of important benefits in many areas, such as dynamic bandwidth allocation, network security and network planning and predictive congestion control and so on. We can identify two categories of predictions: long and short period's predictions. Traffic prediction for long period gives a detailed forecasting of traffic models to evaluate future capacity requirements, and therefore permits for more minute planning and better decisions. Short period prediction (milli-seconds to minutes) is linked to dynamic resource allotment. It can be used to improve Quality of Service (QoS) mechanisms as well as for congestion control and for optimal resource management. It can also be used for routing packets. Several different techniques including time series models, modern data mining techniques, soft computing approaches, and neural networks are used for network traffic analysis and prediction. This paper presents a review of several techniques proposed, used and practiced for network traffic analysis and prediction. The distinctiveness and restrictions of previous researches are discussed and typical features of these network traffic analysis and prediction are also summarized. The remaining paper is organized as follows. A short description about network traffic analysis follows by an extensive review of several available network analysis techniques in section two. Section three reviews various network traffic prediction techniques. In the last section, we give our conclusions.

## 2. Network traffic analysis

Network traffic analysis has become more and more vital and important in present day for monitoring the network traffic. In the past years, administrators were monitoring only a small number of network devices or less than a thousand computers. The network bandwidth was may be just less or 100 Mbps (Megabits per second). Currently, administrators have to deal with higher speed wired network (more than 1Gbps (Gigabits per second)) and various networks such as ATM (Asynchronous Transfer Mode) network and wireless networks. They require additional modern network traffic analysis tools in order to manage network, solve the network problems quickly to avoid network failure, and handle the network security.

Thus, Network traffic analysis presents a number of challenges in recent days. Network is analyzed at different levels viz. at packet level, flow level and network level for security management. Various techniques are being used by researchers for network traffic analysis. A generic framework for network traffic analysis involves preprocessing followed by actual analysis and observations to reveal patterns from the network data. Figure 1 shows three main phases of network traffic analysis. The detailed description of these three phases is presented in subsequent subsections.

### 2.1. Data sets

Testing and evaluating is an important of network traffic analysis. In order to evaluate the effectiveness of all research works using similar standard list is recommended to use standard data set. There are several standard data sets used throughout the recent years. We enlist a few important data sets that are being used by researchers for network traffic analysis.

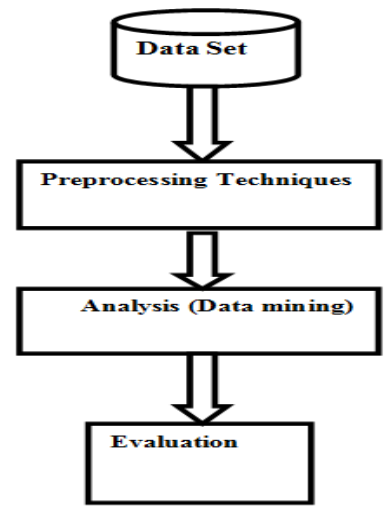

Figure 1. Generic structure of network traffic analysis

1. DARPA data set

KDD cup data has been the most widely used for evaluating of network traffic analysis with respect to intrusion detection. This data set is presented by Stolfo et al. [76]. It is constructed based on the data captured in DARPA IDS evaluation program. The KDD cup data set consists of approximately 4,900,000 training instance and contains 41 features. Moreover, the test data set contains 300,000 instances. The KDD cup contain 24 training and testing attacks and extra 14 types are reported in test data.

2. NSL-KDD data set

The NSL-KDD data set is an updated version of KDD cup data set [77]. The NSL-KDD data set not include redundant instances in training data and duplicate records in testing data. Hence the classifier becomes more accurate. The NSL-KDD is publicly available for researchers and it is improved version of original KDD cup data set.

3. CAIDA data set

This data set contains DoS attacks [78].

4. Waikato data set

It contains internet traffic storage [79].

5. Berkeley Lab data set
It contains internet traffic aarchive [80].
6. ACM SIGCOMM'01 data set
 It contains wireless network traffic [81].
7. DARPA data set
It was a first of network traffic analysis data set with respect to intrusion detection system (IDS). The DARPA data set was released in 1998. The DARPA data set contains two parts: an off-line assessment and a real time assessment that were conducted by MIT Lincoln Laboratory. The DARPA data set contains 38 types of attacks that fall into following four categories: Denial of service (DoS), Remote to User (R2L), User to Root (U2R) and Probing. Seven weeks of network traffic data contain attacks and normal packets for training purpose. The two weeks of network traffic data for testing purpose. These test data contain some attacks that do not exist in the training data. [82] the DARPA include two types of data captured from the network link namely TCP dump and system audit data including Sun Basic Security Module (BSM) audit data from one UNIX Solaris host and file system dump.
8. ISCX data set
It contains normal and DoS attacks [83].

An extensive list of standard data set on network traffic domain is available at [84].

## 2.2. *Preprocessing techniques*

Preprocessing is an important phase use to manipulate real world data into an understandable format. Surely, the real world data have been often incomplete, noisy in specific behavior. In other words, most of data that we wish to analyze from real world by using data mining techniques are incomplete and inconsistent (containing errors, outlier values). Hence, the preprocessing methods are required before applying data mining techniques to improve the quality of the data, thus assisting to enhance the accuracy and efficiency of resulting data mining task. The preprocessing techniques are vital and important in network traffic analysis due to the patterns of network traffic which have different type of format and dimensionality. In next subsections, we provide detailed descriptions of these methods that are used in network traffic analysis.

### 2.2.1. *Discretization method*

Discretization is a preprocessing technique. Discretization is a process of mapping continuous attributes into nominal attributes. The main objective of the discretization process is to discover a set of cut points, which divide the range into small number of intervals. Every cut-point is a real value within the range of the continuous values, which splits the range into two intervals one is greater than the cut-point and the other is less than or equal to the cut-point value. Data discretization techniques can be used to reduce the number of values of continuous attribute by dividing the range of the attribute into small intervals. Interval labels can used to replace actual attributes value. Replacing values of a continuous attribute by a small number of interval labels reduce and simplifies the original data set. Discretization process is an important preprocessing technique for reducing time of network traffic analysis. Discretization methods can be classified into four categories.
(I) Supervised and Unsupervised method
Supervised methods employ the class labels through the discretization process. Unsupervised methods do not utilize information about the class labels and generate discretization methods by sharing the values of the numerical attributes.
(II) Global and Local method
Global methods use entire numerical attributes for the discretization. However, local methods use a subset of instance when obtaining the discretization.
(III) Top-down (splitting) and Bottom-up (merging)
Top-down (splitting) discretization methods begin with long as and value of interval then divide values into smaller intervals at each iteration. However, the bottom-up methods begin with the major number of sub-intervals and combine these sub intervals until achieving optimal number of intervals.
(IV) Direct and Incremental method
Direct methods divide the range of values into equal numbers of intervals and users can determine the number of intervals Incremental methods start with a simple discretization and go during an enhancement procedure till obtaining good discretization then stopping the discretization process.

### 2.2.2. *Feature Selection method.*

Feature selection (FS) is a preprocessing method to be applied before applying data mining techniques. Feature selection used to improve the data mining techniques performance through the removal of redundant or irrelevant attributes. Feature selection methods generate a new set of attributes by selecting only a subset of the original attributes.
 Feature Selection is used mainly to reduce dimensionality of data set for improving network traffic analysis. We present various preprocessing techniques that are being used by researchers before actual analysis of network traffic. We have identified some techniques including principal component analysis, information entropy, rough set theory, feature selection are used frequently for preprocessing network traffic data

**2.3. *Data mining***

Data mining (DM) is used for knowledge-discovery. Data mining plays an important role in analyzing network traffic. Our intension is to present various data mining techniques that are used by researchers for analysis network traffics. We have categorized data mining techniques under four broad categories namely clustering, classification, hybrid and association rules techniques as shown in Figure 2.The detailed description of each technique and its usage is presented

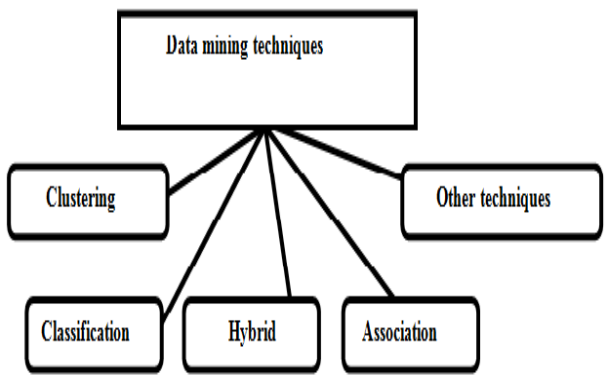

Figure 2. Data mining techniques

2.3.1. *Clustering technique*

Clustering is the process of partitioning data into groups according to certain characteristics of data. Clustering splits data into groups of similar objects. Every group, called cluster, consists of members that are quite similar and members from the various clusters are different from each other. Clustering methods are used for forming groups of network data for network traffic analysis. We identified that several clustering data mining algorithms are used by researchers.

M. Vijayakumar et al. [40] proposed hierarchical clustering data mining algorithm for data stream network traffic. They initially used K-means partition algorithm to form cluster of network traffic data stream. An appropriate cluster is identified and further splitted by using hierarchical techniques to obtained additional clusters. According to authors, this bi-section K-Mean clustering result is better than pure hierarchical clustering method in term of traffic analysis .

Bai Yu, Hong et al. [42] proposes clustering data mining Auto Class tool and K-means algorithms to evaluate the performance impact on wireless mesh networks. Authors collected traffic data from the wireless mesh networks which include both TCP and UDP over IP protocol. They used Auto Class tool and the K-means algorithm to classify and identify session clusters. After clustering log data, the race-driven NS-2 simulations with mixed wavelet model method is presented for analysis of network traffic. A comparison between mixed wavelet (WM) model and traditional Poisson model is presented. They concluded that the mixed WM give better performance to analyze the long-range dependence on wireless mesh networks.

Jeffrey Erman et al. [58] proposed three clustering algorithms, K-means, Auto class and DBSCAN for network traffic classification. Authors collected packets trace from Auckland IV trace and university of Calgary trace with different time interval. They applied K-means, Auto class and DBSCAN algorithms to evaluate the effectiveness of each algorithm for traffic classifications. A comparative analysis between K-means and DBSCAN and Auto class algorithms is presented. From result, they concluded that the Auto Class algorithm performed better for network traffic classification.

Xin Du, Yingjie et al. [70] used K-means clustering data mining technique for intrusion detection. They collected data set from MIB network different time interval. The authors applied information entropy to select most significant attributes from entire set of data set. They selected four attributes destination IP, source IP, destination port, and source port. They inserted various types of attacks into the network data set. They used K-means clustering data mining to determine normal and malicious class. They concluded that the approach has given good performance to select better feature attributes and high accuracy detection rate.

Farhad Soleimanian et al. [68] introduced K-means and Fuzzy K-means data mining clustering approaches for intrusion detection system. They collected data set from KDD cup which has 41 attributes. They categorized the KDD cup data set in to four groups namely fundamental attributes, content attributes, traffic attributes based on time and traffic attributes based on host. They used k-means and fuzzy k-means approaches to identify the type of DoS attacks. A comparative analysis between k-means and fuzzy k-means approaches is presented. They concluded that the fuzzy k-means method achieved slightly better than k-means method in detecting type of denial of service (DoS) attacks.

Miller et al. [66] proposed Den Stream and frequency histogram approaches for detecting anomalous packets based on data stream mining. Authors collected network data from different database DARPA and MCPAD data set. The MCPAD data set contains three types of attacks namely Generic HTTP, Shell-code and Polymorphic attacks. The Den Stream approach treat individual packets as points and are flagged as normal or malicious based on whether these points are normal and outliers. They utilized histogram approach to build histogram model for new packet payload. They used Pearson correlation for computing between two histograms. The sensitivity and false positive rate metrics are applied to measure the performances of algorithms. From the result, they concluded that histogram-based detection algorithm achieved little better performance but required more numbers of features than the clustering-based algorithm.

### 2.3.2. *Classification techniques*

Classification is a form of data analysis which takes each instance of a data set and assigns it to particular class. A classification based network traffic analysis attempts to classify all traffics as either normal or malicious. The challenge of classification is to reduce the number of false positives (detection of normal network traffic as abnormal) and false negatives (detection of malicious network traffic as normal). In this sub section, we present several classification data mining algorithms that are being used by researchers for network traffic analysis.

(a) Support vector machine (SVM) approach

Support vector machine (SVM) is a supervised learning method used for classification and regression. The Support Vector Machine was created by Vapnik [64] support Vector Machine use to train and test large amounts of high dimensional data. The SVM is very costly in terms of time and memory consumption. A number of current studies have reported that the SVM results of giving higher performance with respect to classification accuracy than the other classification approaches.

Rung-Ching Chen [43] proposed SVM with Rough Set Theory (RST) for intrusion detection. They used rough set theory to reduce dimensionality of a DARPA database. Furthermore, the features are selected by using rough set theory. Then, they applied SVM approach to train and test respectively. A comparative analysis for results between SVM method with entropy, SVM with original data set and SVM with RST is presented. They concluded that the SVM with RST achieved better performance.

Ghanshyam Prasad Dubey et al. [51] proposed RST and incremental SVM approaches to detect intrusion. Authors experimented with KDD cup dataset. The selection of significant attributes from network traffic data set is completed using RST method then later processed by SVM approach for training and testing. From data analysis, a comparative between incremental SVM and non-incremental SVM is presented. According to authors, the incremental SVM approach increased performance for intrusion detection. They noted the RST and increment SVM approaches are effective to decrease the space density of data.

Shailendra Kumar et al. [64] introduced rough set theory and support vector machine for dimensionality reduction in intrusion detection. They experimented with KDD cup data set. They applied rough set theory to select most significant attributes from KDD cup data set. They experimented with original 41 attributes data set and with reduced data set (29 and 6 attributes). A comparative analysis between SVM with original 41 data set and reduction data set is presented. They concluded proposed algorithm is very reliable for intrusion detection.

Heba F. Eid et al. [73] analyzed intrusion detection system using Support Vector Machines with Principal Component Analysis approaches. They tested their proposed model on NSL-KDD data set. The PCA approach used to reduce the number of features in order to decrease the complexity of the system. The features that has been selected by PCA later processed by SVM approach for classification. Their results showed that the proposed system is capable to speed up the process of intrusion detection and to reduce the memory space and CPU time cost.

(b) Neural network approaches

Neural network consists of a collection of processing neurons that are extremely interrelated and transform a set of inputs to a set of required outputs. The result of the transformation is recognized by the characteristics of the neurons and the weights connected with the correlation among them. By enhancing the connections between the nodes network is able to obtain required outputs. We show several neural network techniques that are being employed by researchers for network traffic analysis.

Khaled Al-Nafjan et al.[69] presented a modular neural network (MNN) for intrusion detection. Authors collected data set from KDD cup data set which contains 41 attributes. They obtained most significant attributes from entire set of network traffic data set by using PCA method. They applied MNN to detect network intrusion. From analysis of result, they communicated that MNN with PCA approach has reduced RMSE to 0.1.

Shilpa Lakhina et al.[53] proposed principal component analysis neural network algorithm (PCANNA) for identifying any kind of new attacks. The PCA approach used to reduce the dimensionality of data set for improving the accuracy. Test and comparison are done on NSL KDD data set. Furthermore, they selected 8 attributes from entire set of 41 attributes. Results of classification using hybrid model with 41 attribute and 8 attribute are compared and authors concluded the classification followed after PCA preprocessing gives better performance.

(c) Decision tree approaches

The decision tree (DT) is powerful and popular supervised machine learning for decision-making and classification problems. The decision tree is employed in the number of real life applications such as medical diagnosis, weather prediction, credit approval, and intrusion detection etc. DT can use large volume of data set with many attributes due the tree size. It is independent of the data set size. A decision tree consists of nodes, leaves, and edges. Each node is tagged with on attribute against all attributes by which the data is to be partitioned. Every node has a number of edges according to possible values of the attribute. Leaves are tagged with a decision value for classification of the data. Every decision tree corresponds to a rule set, which classify data according to the attributes of data set. The DT building algorithms may in first, create the tree and then prune it for more efficient classification. With pruning technique, portions of the tree may be removed or combined to reduce the overall size of tree the time and memory consumption of decision tree depends on the size of data set [85].

Xu Tian et al. [44] proposed two stream mining algorithms namely Data Stream based traffic (DSTC) and Very Fast Decision Tree (VFDT) for online network traffic classification. They collected data set from especially peer to peer like Bit Torrent, PP Live from several applications at different time interval varying size of data. The feature selection method is used to select most significant attributes from set of network traffic data set. A comparative analysis among VFDT and C4.5 and Bayes Net is presented. They communicated that the VFDT is more accurate, less memory consuming and update fast.

Jiong Zhang, et al. [62] used unsupervised random forest algorithm to detect outliers in data set of network traffic. The outlier detection is used for network intrusion purpose. The authors experimented with the KDD cup dataset. A comparative analysis between supervised and unsupervised algorithms is presented. Authors concluded that their unsupervised outlier detection approach resulted less false positive cases as compare to neural network (NN) and SVM supervised classification algorithms.

Sandhya Peddabachigari et al. [50] proposed decision tree data mining techniques for investigating and evaluating intrusion detection. The authors collected data set from DARPA which has different type of attacks. They divide the five classes into normal, Probe, DoS, user to root (U2R) and remote to user (R2L) classes. A comparative analysis between decision tree and SVM is presented. They communicated that the decision tree provides better accuracy than SVM on Probe, U2R, R2L and normal classes. The SVM gave better result than decision tree on DoS class. They observed that the decision tree is more efficient for detecting the intrusion.

Arya, Mishra [49] proposed four data mining approaches namely J48, Random Tree, Random Forest and Boosting to improve performance of internet traffic. The authors experimented with the NLANR data set. They selected nine applications (namely HTTP, SMTP, DNS, SOCKS, IRC, FTP Control, POP3, and LIMWIRE) from network traffic. They combined the five classifiers into multilevel classifiers to enhance performance of classification. The accuracy, recall and precision metrics are used to determine the effectiveness of multilevel classifiers. A comparative analysis between multilevel classifiers and non multilevel classifiers is presented. They concluded that the multilevel is more accurate to classify internet traffic. The drawback of multilevel classifier is that it requires more time to build the model.

(d) Statistical approaches

Bayesian classification is based on bayes theorem. A bayesian classifier is based on the idea that, if an agent knows the class, it can predict the values (instances) of the other attributes, if it does not know the class, Bayesian rules can be used to predict the class given some of the feature values. We survey statistical algorithms that are employed by researchers for network traffic analysis [85].

G.kalyani et al. [55] proposed several types of classification data mining techniques namely Naive Bayes, J48, OneR, projective adaptive resonance theory (PART) and RBF network algorithm for intrusion detection. Authors collected data set from KDD cup which has 24 types of attacks. A comparative analysis between these approaches is presented. They noted that the PART algorithm achieved better result than other techniques for intrusion detection.

Jaspreet Kaur et al. [41] introduced five machine learning algorithms namely Naive Bayes, Radial Basis Function (RBF), multilayer perceptron (MLP), C4.5 and Bayes Net to detect intrusion. They collected data set from educational and non-educational websites by using wire shack software. They used Recall, Precision and Accuracy metrics to measure the performance of five approaches. A comparative analysis between five approaches is presented. They noted that the Bayes Net performed better for detecting intrusion.

Yogendra et al. [74] study four different data mining approaches (J48, BayesNet, OnerR and NB) for detecting intrusion and compared their relative performance. They concluded that the J48 classifier achieved better result with high detection rate and low false positive and cost. Yogendra Kumar Jain et al. [59] proposed naive bayes and Bayes Net supervised learning algorithm for intrusion detection. They collected data set from KDD cup data set. The dimensionality of network traffic data set is reduced using information entropy method. They compared their results with RC [86]. They shared that their results are better than RC Staudemeyer research paper result. A comparative result between NB and Bayes Net is presented. They concluded that BayesNet with an accuracy rate of approximately 99% performs much better at intrusion detection than NB.

Neethu [67] presented Naive Bayes classifier with Principal Component Analysis (PCA) algorithms for intrusion detection. They collected data set from KDD cup. The dimensionality reduction of input data is obtained by Principal Component Analysis. A comparative analysis between original data and reduction data is presented. They observed that the naive Bayesian network approach achieved better result in terms of false positives in network traffic and also, it require less time and low cost for building model.

Rupali Datti et al. [71] proposed linear discriminate analysis algorithm to extract features for detecting intrusions and Back Propagation algorithm is used for classification of attacks. They tested their models on NSL-KDD data set which is improved version of KDD cup data set. The experimental results showed that the model gives better results for selecting most significant features from entire data set.

### 2.3.3. *Hybrid models*

The hybrid models are a combination of two or more approaches for analysis of network traffic. The hybrid model achieved good results in the analysis of network traffic. We present various hybrid model techniques that are investigated by researchers for network traffic analysis.

Yacine Bouzida et al. [34] introduced two machine learning algorithms namely support vector machine (SVM) and decision tree (DT) to detect intrusion in term of network traffic analysis. Authors collected data set from KDD cup which contains 24 attacks. They applied PCA to reduce the dimensionality of data set. A comparative analysis results between DT with PCA and DT without PCA are presented. They concluded that the DT with PCA achieved little better than DT without PCA. Moreover, they compared between SVM with PCA and SVM without PCA. They shared that SVM with PCA resulted good performance.

Venkata Suneetha et al. [35] presented hybrid model using poly kernel SVM and K-means algorithm for intrusion detection model. They experimented with the KDD cup data set. They used Information Gain and Triangle Area based KNN for reducing the dimensionality of data set. The hybrid model used to analyze normal and malicious packets. They shared that their model achieved high accuracy detection rate and less error.

Jashan Koshal et al. [45] proposed hybrid model for developing the intrusion detection system by combining C4.5 decision tree and Support Vector Machine (SVM) approaches. They collected data set from KDD cup. The preprocessing of data reduced the dimensionality of entire network traffic data set using feature selection method. They selected 12 attributes among from 41 attributes. They applied hybrid model to detect normal and malicious packets. A comparative analysis between single approaches and hybrid approaches are presented. They shared that the hybrid approach is more efficient than single model for detecting intrusion.

Xiang He et al. [11] introduced rough set theory and neighborhood rough set theory approaches for intrusion detection. Authors experimented with KDD cup data set which contains 41 attributes and 24 attacks. They applied global discretization method to convert data set from continuous attributes in to discrete attributes. After data preprocessing, they used rough set theory approach to reduce dimensionality of original data set. The detection rate metric is presented to measure the performance of rough set theory approach. A comparative analysis of results between rough set theory and neighborhood rough set theory approaches is presented. They communicated that the neighborhood rough set theory approach achieved high detection rate.

R.Lath et al. [48] proposed k-mean cluster, k-nearest neighbour and SVM classification approaches for intrusion detection. They experimented on KDD cup. They used statistical normalization method to select most significant features from KDD cup data set. A comparative analysis between k-mean cluster, k-nearest neighbour and SVM approaches is presented. They concluded that k-nearest neighbour is more accurate to detect normal and malicious packets but consumed more time to build model.

Srinivas Mukkamala et al. [52] presented support vector machine (SVM) and neural network to evaluate for intrusion detection. They collected data set from DARPA. The selection of most significant attributes from network traffic data set is obtained by SVM approach. They selected 13 attributes that are most significant out of 41 attributes. They applied SVM and neural network approaches for classifying normal behavior and attack patterns. They used positive rate, accuracy metrics at the training and testing time to measure performance of their approaches. The result of analysis using SVM and neural network approaches with original 41 attributes and with reduced 13 attributes are compared. The authors concluded that the SVM approach show little better than neural network approach.

Warusia Yassin et al. [57] introduced an integrated machine learning K-means (KM) clustering and Naive Bayes(NB) classifier approaches to detect intrusion. They evaluated their techniques with real network traffic data set, collected over several days from ISCX network traffic data set. A comparative experimental result between individual Naive Bayes classifier (NB) approach and an integrating KM+NB approaches is presented. They shared that an integrated KM+NB approach achieved higher accuracy.

Sadek et al. [72] presented a hybrid namely algorithm Neural Network with Indicator Variable using Rough Set for attribute reduction (NNIV-RS). It is used to reduce the amount of computer resources like memory and CPU time required to detect attack. Rough Set Theory is used to selected features from data set. They used Indicator Variable for representing data set in more efficient way. Neural network approach is presented to determine normal and malicious packet from network traffic. Tests and comparison were done on NSL-KDD data set. The experimental results showed that the neural network gave better results.

### 2.3.4. Association rules data mining techniques

The association rule considers each attribute/value pair as an item. Collection of items referred as an item set in a single network request. Association rules are used to identify pattern or relation among the attribute of data base. Association rules are important for analysis of network traffic. We survey different association rules algorithms that are employed by researchers for network traffic analysis.

Zulaiha Ali Othman et al. [37] proposed Fuzzy Apriori, FP-growth and Apriori association rule data mining to detect normal and malicious packets from network traffic. The authors collected data from University of Kebangsaan Malaysia (UKM) at different time intervals by using Wireshark tool. They developed a new tool for intrusion detection called as Nasser tool. The Nasser tool consists of four components namely, preprocessing, option setting, association techniques and analysis components. The preprocessing component is used to transform and select most significant features from network data set. The option setting use for setting time interval for which the analysis shall be carried out. Association rule component includes different algorithms namely Fuzzy Apriori, FP-growth and Apriori to determine normal and malicious class. The analysis components used to analyze the result. A comparative analysis between Fuzzy Apriori, FP-growth and Apriori association rule approach is presented. They shared that the Fuzzy Appriori approach is more accurate while FP-growth approach is faster. Further, they concluded that the their tool achieve better performance for intrusion detection.

Weisong. He, Guangminhu et al. [46] proposed time series association data mining to detect intrusion in network traffic analysis. They collected data set from backbone network Abilene, connecting over 200 universities. They compute entropy to reduce uncertain attributes from entire data set. The PCA approach based on subspace method used to select most significant attributes from the data set. They applied piecewise aggregate approximation (PAA) and symbolic aggregate approximation (SAA) methods to represent time series. They applied association rules data mining for detecting intrusion.

He, Guangmin Hu et al. [47] introduced multivariate time series association rule data mining to analyze network traffic for intrusion detection. They collected data set from real network (intenet2 backbone network) over 20 universities server. The proportion-based analysis method used to filter TCP flag with time series into destination IP, source IP, destination port, and source port. They used PAA method to represent time series on TCP flag and convert to discrete symbolic elements. Furthermore, they applied SAX method for the analysis of discrete symbols. The authors presented multivariate time series association rule data mining to determine the normal and malicious packets. They concluded that their approach achieved good performance for intrusion detection.

### 2.3.5. Other techniques

Other techniques are defined as various methods for analysis of network traffic. These techniques are applied on different algorithms for network traffic analysis. We present several algorithms that are used by researchers.

Yingjie Zhou et al. [26] proposed time-series Graph Mining for detecting anomalous packets from network traffic. They collected data set from Abilene backbone internet at different time intervals. The four attribute from data set namely source IP, destination IP, source port, destination port are selected from header of packets. They presented this value of attributes and relationship between them by using time series graph. The weight coefficients function used to compute the relationship between attributes for detecting malicious class. A comparative analysis between Graph Mining and Continuous Wavelet Transform-based (CWT) approaches are presented. They concluded that the graph mining is very accurate to detect distribution denial of service (DDoS) attacks out of different type of attacks.

Zhou Mingqianet al. [60] proposed Local Deviation Coefficient Graph Based (LDCGB) algorithm for intrusion detection. Authors collected data set from Massachusetts Institute of Technology which has 24 types of attacks and divide attacks into four types Denial of Service, Remote to User, User to Root and Probing attacks. The graph based clustering algorithm is proposed for partition data set into a number of clusters. Further, they applied Local deviation coefficient method regroup in one of the clusters. They used detection rate and positive rate metrics to measure the effectiveness of LDCGB algorithm for detecting malicious class. A comparative experimental result between K-means and fuzzy C means (FCM) approaches and LDCGB approach is presented. They shared that the FCM approach gives better performance for intrusion detection than K-mean approach and FCM approach.

Nikita Gupta et al. [27] presented rough set theory to reduce dimensionality as well as classification. Authors collected network traffic data set from NSL-KDD database. They proposed rough set theory to extract the relevant attributes from entire data set and for classification. They applied accuracy and sensitivity metrics to measure the performance of rough set theory based intrusion detection. They concluded that the rough set theory approach achieved high accuracy to reduce dimensions of attribute set and also to detect intrusion.

P.Gifty et al. [38] Presented fisher linear discriminate analysis (FLDA) classification data mining to detect malicious packets at time of accessing network. Authors experimented with KDD cup data set. They applied correlation function based feature selection method to reduce dimensionality of data set from KDD cup benchmark. The fisher linear discriminate analysis used to classify normal and malicious classes. According to authors, the FLDA approach is more efficient to detect R2L and U2R attacks out of four attacks.

Pooja et al. [75] analyzed NSL-KDD data set to determine most significant features set of DoS attack from original data set. They divided 41 original features into three categories (basic features, traffic features and content based features). They concluded that when more relevant features are available for classifications then it improve performance of intrusion detection. Moreover, the accuracy increased and the time of classification is reduced.

### 2.4 Evaluation metrics

In data mining techniques, many different metrics are used to investigate the data mining techniques. The detection rate, false positive rate, accuracy and time cost metrics are employed for measuring the performance of classifier for different data set. A number of metrics exist to express predictive accuracy. The metrics used using confusion matrix. Each metric is defined as below

(a) True Negatives (TN)

Total numbers of normal packets correctly classified.

(b)True Positives (TP)

Total numbers of malicious packets correctly classified.

(c) False Negatives (FN)

False Negatives is total numbers of malicious packets incorrectly classified as normal packets.

(d) False Positives (FP)

False positive is Total numbers of normal packets incorrectly classified as malicious packets.

(e) Detection rate (DR)

It is the ratio of total numbers of attacks detected divided by total numbers of false positive plus total number of true negative.

(f) Precision rate (PR)

It is the ratio of total numbers of TP divided by total number of TP plus total number of FP.

(g) Recall rate (RR)

It is ratio of total numbers of TP divided by total number of TP plus total number of FN.

(h) Overall rate (OR)

It is ratio of total numbers of TP pulse total number of TN divided by total number of TP plus total number of FP plus total number of plus total number of TN.

(i) Sensitivity

It is the ratio of total numbers of TP divided by total number of FP.

(j) Specificity

It is the ratio of total numbers of TN divided by total number of FN.

(k) Accuracy

It is the ratio of total numbers of TP plus total numbers of TN divided by total number of FP plus total number of FN.

(l) Percentage of successful prediction (PSP)

It is the ratio of total numbers of successful instances classified divided by the total numbers of actual instance.

Table 1. Network traffic analysis techniques

| Authors, year | Preprocessing | Techniques | Data set | Purpose | Evaluation metrics |
|---|---|---|---|---|---|
| M. Vijayakumar et al. 2010 | K-means | Hierarchical clustering stream network traffic | From internet service provider | Stream network traffic analysis | Accuracy |
| Bai Yu, Hong Fein et al. 2008 | Not mentioned | K-means | NS2 simulation over TCP and UDP protocol | Avoid congestion of mesh wireless network in long-rang dependence | Accuracy |
| Jeffrey Erman et al. 2006 | Not mentioned | K-means, Auto class and DBSCAN | Auckland IV and university of Calgary | For Network traffic classification | Accuracy |
| Jiong Zhang et al.2006 | Extracting the attributes from data set by using by feature selection method | Random forests | KDD cup | For intrusion detection | Accuracy, FP |
| Xin Du, Yingjie et al.2008 | Information entropy | K-means | MIB network | For intrusion detection | Accuracy |
| Farhad Soleimanian et al.2010 | They divide data set into four categories namely fundamental, host, network and based on time attributes | K-means and Fuzzy K-means | KDD cup | For detecting DoS attack | DR, Accuracy |
| Miller, W.Deritrik et al. 2011 | Histogram payload packets | Den Stream and frequency histogram | DARPA and MCPAD | For intrusion detection | Accuracy, DR, FP |
| Rung-Ching Chen et al. al.2009 | Rough Set Theory | Vector Machine (SVM) | DARPA dataset | For intrusion detection | DR, Accuracy |
| Ghanshyam Prasad et al. 2011 | Rough Set Theory | Increment SVM | DARPA | For intrusion detection | Accuracy |
| Shailendra Kumar et al. 2011 | Rough set theory | Support vector machine | KDD cup | Intrusion detection | Accuracy, FP |
| Shilpa lakhina et al. 2010 | PCA | Neural network | DARPA | Intrusion detection | Accuracy |
| R.Lath et al. 2012 | Statistical normalization method | K-means, SVM, K-nearest Neigh bour | DARPA | Intrusion detection | FP, Accuracy |
| Neethu 2008 | PCA | Naïve Bayes | KDD cup | Reduce false positive for intrusion detection | Accuracy, DR, FP |
| Yogendra Kumar et al. 2011 | Information entropy | NB and Bayes Net | KDD cup | For increase detection rate of intrusion detection | Accuracy, Recall, Precision, F-Measure |
| Khaled Al-Nafjan et al. 2012 | PCA | Modular neural network | KDD cup | Reduce false positive rate of intrusion detection | RMSE |
| Yingjie Zhou et al. 2009 | Not mentioned | Graph Mining | Abilene data set | Intrusion detection | Accuracy |
| Xu Tian, Qiong Sun et al. 2008 | Feature selection | Fast Decision Tree | peer-peer Bit Torrent, PP Live | Stream network traffic | Accuracy, DR |
| Zhou Mingqiang et al. 2012 | Not mentioned | Local Deviation Coefficient Graph Based | Massachusetts Institute of Technology dataset | Intrusion detection | Accuracy, DR |
| Arya and Mishra et al.2011 | Feature selection | J48, Random Tree, Random Forest and Boosting classifiers | NLANR dataset | Analysis internet traffic | Precision, Recall, Accuracy |
| Yacine Bouzida et al. 2004 | PCA | SVM and decision tree | KDD cup | Intrusion detection | PSP |
| Rupali Datti et al. 2010 | Not mentioned | linear discriminate analysis | NSL-KDD | For extracting feature from data set for intrusion detection | Accuracy |
| Nikita Gupta et al. 2013 | Rough set theory | Rough set theory | NSL-KDD | Intrusion detection | Accuracy, Sensitivity |
| Jashan Koshal et al .2012 | Features selection | Decision tree and Support Vector Machine | KDD cup | Intrusion detection | FP,FN,TP,TN |
| Venkata Suneetha et al. 2005 | Not mentioned | Triangle area based on KNN and poly kernel SVM and K-means | KDD cup | Intrusion detection | Accuracy, FP |
| Srinivas Mukkamala et al. 2010 | SVM | Support vector machine and neural network | DARPA dataset | Intrusion detection | Accuracy |
| Warusia Yassin et al. 2013 | Not mentioned | An integrated machine learning K-means clustering and Naïve Bayes classifier | From server of ISCX | Detection DDoS attack | TP, TN, FP, FN |
| Ming-Xiang He et. 2013 | Rough set theory and discretization method | Neighborhood rough set | KDD cup | Intrusion detection | DR, FP |
| G.kalyani et al. 2012 | Feature selection | Naïve Bayes, J48, Oner, PART and RBF network algorithm | KDD cup | Intrusion detection | RAE, RMSE, RRSE, TP, FP |
| Jaspreet Kaur et al. 2012 | Not mentioned | Naïve Bayes, RBF, MLP and Bayes Net | Wireshark software | Intrusion detection | Accuracy, Recall, Precision |
| Zulaiha Ali et al. 2011 | Discretization | Fuzzy Apriori | University Kebangsaan Malaysia (UKM) | Intrusion detection | Accuracy, DR |
| He, Guangmin Hu et al. 2009 | PAA,SAX method | Time series association data mining | from Abilene (intenet-2 backbone | For detection normal and malicious packets | Accuracy, DR |

| | | | network) over 20 universities servers | | |
|---|---|---|---|---|---|
| Weisong, Heisong He et al. 2008 | Information entropy | Association rule mining | from Abilene (intenet-2 backbone | Minimize false positive and false negative | DR,FP |
| Rowayda.Sadek et al. 2009 | Rough set theory | Neural Network with Indicator Variable using | NSL-KDD | Reduce dimensionality of data set for improve intrusion detection | DR,FP |
| Sandhya peddabachigari et al. | Not mentioned | DT | DARPA | For intrusion detection | Accuracy |
| Heba F. Eid et al 2010 | PCA | SVM | NSL-KDD | For improve time and memory cast for detection intrusion | Accuracy |
| Yogendra et al. 2012 | Feature selection | J48, BayesNet, OnerR and NB | KDD cup | For intrusion detection | RAE,RMEE, Accuracy |
| Pooja et al.2013 | Basic feature, traffic features and content based features | Decision tree | NSL-KDD | For detecting DDoS attack | Accuracy |
| P Gifty et al. 2012 | Correlation function | FLDA | KDD cup | For intrusion detection | FP, Accuracy, DR |

## 3. Network traffic prediction

Network traffic prediction is an important issue that has received much interest recently from computer network community. The network traffic prediction is one of the typical issues useful for monitoring network, network security, avoid congestion and increase speed of networks. Different techniques are used by researchers for network traffic prediction. We have categorized these techniques under four broad categories namely linear time series model, nonlinear time series model, hybrid model and decompose model and. Figure.3 gives an overall idea of these four categories under which various techniques of network traffic prediction are categorized. The detailed description of each phase is presented in following sub section.

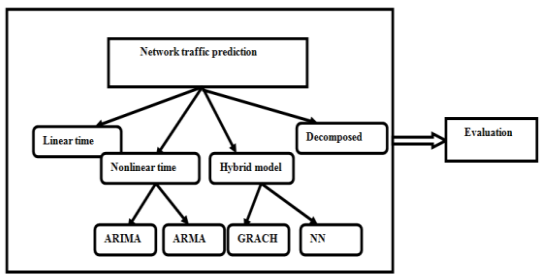

Fig.3. Network traffic prediction technique

### 3.1. Linear time Series technique

Linear Time Series techniques are covariance structure in the time series. There are two popular sub-groups of linear time series models: Auto Regressive (AR) and the Moving Average (MA) models, which can be combined to make the auto regressive moving average models. Linear time series is traditional technique in network traffic prediction. We present various linear time series techniques that are being used by researchers for network traffic prediction

#### 3.1.1. Autoregressive Moving Average (ARMA)

The ARMA model is combined from both AR and MA to obtain an accurate model. Autoregressive Moving Average is statistical model use with time series model for prediction. The ARMA model applied to well-behaved time series data. ARMA model are considered in adequate for predicting data that integrate random. The ARMA linear time series model is very significant to predict network traffic. We discuss ARMA algorithms that are employed by researchers for network traffic predictions as follows.

N. K. Hong, et al. [14] introduced ARMA time series model for predicting large file transfer of network traffic. Authors experimented with file transfer protocol (FTP) data set. The network protocol analyzer (wireshark) is applied to capture and collect packets at different time intervals. They applied necessary transformation (log return) for computing the number of packet while capturing data. They transferred 1GB file and 10 MB file through network. Then applied ARMA approach for forecasting different files at different time interval. They concluded that transfer of small file is better than large file. And also, they recommended breaking a large file into several small size files. Transmitting smaller files lead to use of little bandwidth. They suggested that the most appropriate model to be used is the ARMA different size of file transfer.

Poo kuanHo et al. [16] proposed a time series model with ARMA approach to predict network traffic pattern of bit torrent application. They evaluated ARMA technique, collected 6 sets of real data at a period of six days from bit Torrent P2PN (point to point network) by using wireshark software. They developed a new simulation tool called as simple network prediction studio (SNPS). This simulation used ARMA model to predict cyclical and seasonal bit torrent network traffic patterns. The mean square errors (MSE) metric applied to measure and evaluate performance of ARMA model on cyclical and seasonal patterns. A comparative prediction result of ARMA model at cyclical and seasonal patterns of bit torrent is presented. They shared that the ARMA approach achieved better prediction in cyclical pattern.

Nayera Sadek et al. [20] proposed time series model with K-factor ARMA model for predicting multi scale high speed network traffic. They collected dataset from different network applications such as MPEG, VIDEO, JPEG, INTERNT, and ETHERNT. They used mean absolute error (MAE) and error ratio(ER) functions to measure prediction performance of K-factor ARMA. The K-Factor ARMA model predicted different types of network traffic in time domain and frequency domain. They compared performance between K-factor ARMA model and traditional autoregressive (AR). They concluded that their model achieved better performance for prediction of network traffic.

Yanhua Yu et al. [19] proposed accumulative predicting model (APM) method with time series model to predict mobile network traffic. They collected dataset from certain mobile network of Heilongjian province in china. They applied means absolute percentage error (MAPE) to measure and evaluate prediction of APM technique. The autocorrelation function is used to represent time series of traffic data. A comparative prediction result between APM approach and ARIMA approach is proposed. They concluded that the APM is more efficient and effective for network traffic prediction.

### 3.1.2. *Autoregressive Integrated Moving Average Model(ARIMA)*

This Model is an integrated ARMA model. ARIMA play an important role with time series model to predict network traffic. ARIMA model is mostly used for forecasting network traffic. We present ARIMA algorithms that are used by researchers for network traffic predictions as follows.

Yanhua Yu et al. [25] proposed ARIMA linear time series model for prediction of network traffic. Authors experimented with mobile of network from Heilongjiang chain at different time interval. They used MAPE metric to measure and evaluate prediction of ARIMA approach. They applied correlation coefficient function to represent time series. According to authors, the ARIMA approach achieved high prediction.

Yantai Shu et al. [29] described linear ARIMA model for modeling and predicting wireless network traffic. The authors used GSM wireless mobile real data from Tianjin at china mobile of different time scales. They applied minimum mean square error (MMSE) function to measure and determine prediction performance of ARIMA model. A comparative prediction between actual data and prediction data is presented. They concluded that the ARIMA technique obtained better prediction. They observed relative error between prediction values and original values are 0.02.

Bozidar vujic, hao chen et al. [31] presented SARIMA linear time series model with data mining model (k-means algorithm) for predicting network traffic. They collected network log data from a deployed public safety trunked radio network with different time intervals during three months duration. They preprocessed data to extract most significant attributes from entire set of attributes. Furthermore, the K-means approach used to cluster data into three groups based on their calling behavior. They used NMS metric to evaluate prediction of SARIMA model. They observed that the SARIMA model with k-means cluster obtained better prediction.

Aqil Burney S.M et al. [28] proposed SARIMA approach based on wavelet filter to reduce noise which present in network traffic. They applied Daubechies db4 wavelet filter method to reduce noise from traffic data by decomposing the original signal. Further, they used wavelet multi resolution method for compression and reconstruction of decomposed signal into various small signals. Later, they fitted SARIMA approach to predict short-range network traffic. They collected real data from university of Karachis High speed Fiber optics Local Area network with wireless computing. From experimental result, they noted that the SARIMA model based on wavelet approach improved the speed and efficiency by reducing noise of network traffic.

Huda M.A. Hag et al. [13] introduced new model Adjusted Autoregressive Integrated Moving Average (AARIMA) for modeling network and to predict long-range dependent internet traffic. Authors collected data set from Start ware MPEG and Bell core data set. The Box-Jenkins method is used to represent time series for building AARIMA model. They used mean absolute error to measure the prediction performance of AARIMA model. A comparative accuracy of Hurst (H) parameter between ARIMA model and AARIMA model is presented. They concluded that the value of H parameter given by AARIMA model is more accurate than ARIMA model.

Shen Fu-ke, Zhang Wei, et al. [5] described engineering prediction time series (EPTS) approach to predict network traffic. Authors experimented with ECNU data set with different time intervals. They configured ECNU by setting rate-limiting on interface at the edge of network to limit into or out of network. They decomposed time series into linear trend, periodic, random components to evaluate the EPTS approach to predict internet traffic. From experiments, they compared between EPTS with rate-limiting and EPTS without rate-limiting. They shared that the EPTS with rate-limiting give more efficient network traffic prediction.

## 3.2. *Nonlinear time series techniques*

Nonlinear time series are generated by non-linear dynamic equations. They exhibit feature that can't be modeled by linear processes such as time-change variance, asymmetric cycle, higher-moment structures, thresholds and breaks. This model used to predict network traffic with different techniques such as neural network, Fuzzy logic. We display various non-linear time series techniques that are being used by researchers for network traffic prediction.

### 3.2.1. *GARCH technique*

Nikkei C.anand, Caterinal et al. [4] proposed nonlinear time series Generalized Auto Regressive Conditional Heteroskedasticity (GARCH) model to capture the bursty nature of Internet traffic. They evaluated their techniques using data from Abilene backbone internet for different time scales. They used one step prediction method; wherever prediction is recursively is completed to get following prediction values at suitable time. GARCH and ARCH metrics are used to measure the performance of their techniques. A comparative prediction between GARCH approach and ARIMA approach is presented. They shared that the GARCH approach gives better prediction.

### 3.2.2. *Neural Network techniques*

Neural network consists of functions called neurons. These neurons have connections to obtain the inputs and pass the output to other neurons. Every connection has a weight associated with it. This weight determines performance of the NN. These weights are learned through the training stage. The neural network for network traffic has been investigated by number of researchers. Neural networks provide a solution to the problem of network traffic prediction by applying numbers of techniques. Neural networks for predicting network traffic were first introduced as an alternative to statistical techniques in network traffic. We show several neural network techniques that are being employed by researchers for network traffic prediction. Edmund S.Yu, et al. [7] used time series neural network to predict noise of video of network traffic. They collected data set from B-ISDN. They used autocorrelation function to represent time series. The mean square error metric used to measure prediction performance of neural traffic. From their network prediction results, a comparative prediction between neural network approach and Box-Jenkins approach is presented. They shared that NN approach is more effective and accurate than traditional linear time series.

Dong Churl park et al. [1] presented non linear time series model with dynamic bilinear recurrent neural network (D-BLRNN) techniques to predict network traffic. They applied D-BLRNN approach to improve the bilinear recurrent neural network (BRNN) approach for network traffic prediction. They have collected real data at Bell core for different duration's time slots. The Mean square error (MSE) metric used to measure prediction performance of D-BLRNN. A comparison between D-BLRNN and BRNN techniques is presented. They communicated that the D-BLRNN gives better prediction than BRNN approach.

Samira Chabaa1, Abdelouhab et al. [10] presented neural network based on multi-layer perception (MLP) to analyze and predict the internet traffic over IP network. They collected network data set from backbone internet. They used relative error means absolute percentage (RMSEP) to measure prediction performance of neural network technique. A comparative prediction between MLP approach and resilient back propagation (RP) approach is displayed. They communicated that the MLP approach achieved better prediction.

Wang Junsong, Wang Jiukun et al. [12] proposed nonlinear time series model Elman Neural Network (ENN) for internet traffic prediction. The mean square error (M.SE) and normalized mean square error (NMSE) metrics are used to evaluate prediction of ENN approach. A comparative prediction result between an Elman neural network, Wavelet and ARIMA model are presented. They concluded that the ENN approach is more feasible and efficient to predict internet traffic.

Samira Chabaa et al. [61] proposed adaptive neural fuzzy inference system (ANFIS) with time series for modeling and predicting internet traffic. Authors collected data from backbone internet over TCP/IP protocol. They used root mean square error (RMSE) and average absolute relative error (AARE) to evaluate the prediction of ANFIS model. They noted that the ANFIS model is better in prediction of internet traffic.

Paulo Cortez, Miguel Rio et al. [33] proposed Neural Network Ensemble (NNE) ARIMA and holt-winter approaches with time series model for predicting internet traffic. Authors collected data from internet service provider based on TCP/IP protocol. They applied their techniques to data obtained at different intervals of time such as five minute, one hour and one day. They used MAPE metric to measure prediction performance of their techniques. A comparative prediction result between NNE, ARIMA and Holt-Winter methods is presented. They shared that the Holt-Winter approach is better to predict network at one day time scale but the ENN and ARIMA approaches achieved better prediction at five minute and one hour time interval but ARMA is very complex than NN. And also they noted that the NN is more accurate overall.

Wang Peng, Liu et al. [22] used Propagation Wavelet Neural Network (PWNN) technique for improving drawback of Propagation Neural Network (PNN) technique in network traffic prediction. They collected network data set from real link for time of fifteen days. They applied Propagation neural network and BPWNN to perform one-step and multi-step network traffic prediction. They used MSE metric to evaluate prediction performance of their technique. A comparative prediction result between the PWNN and BPNN approaches are presented. They noted that in one-step case both approaches perform good prediction but in the multi-step prediction the PWNN is more accurate to predict network traffic.

Hong Zhao [24] presented fast wavelet transform based Least Mean Kurtosis (LMK) method to predict self similar network for reducing complex temporal correlation in short range dependencies in the wavelet domain. They applied Haar algorithm to decompose log data into wavelet coefficients. LMK algorithm is used to predict wavelet coefficients. They used least mean square (LMS) to measure the prediction performance of LMK algorithm. According to authors, the LMK algorithm achieved better prediction result with less computational complexity.

### 3.3. Hybrid model techniques

The hybrid model is a combination of two or more models. The hybrid model is very accurate in predicting network traffic. The combination of linear and nonlinear models is called hybrid model. It gives good results in the prediction and analysis of network traffic. We present various hybrid model techniques that are explored by researchers for network traffic prediction.

S.M.Aqil Burney et al. [15] applied conditional mean and conditional variance model (ARIMA with GARCH) on HTTP request series. Furthermore, they used autocorrelation function to represent time series. They collected data set from department of computer science of university of Calgary with different time scale. A comparative prediction between linear time series SARIMA and non-linear time series SARIMA/ GARCH model is presented. They noted that non-linear time series SARIMA/ GARCH model resulted more accuracy to predict network traffic.

BoZhou, Dan He et al. [32] combined a linear ARIMA and non- linear GARCH time series models. This model was used mainly to predict long-range dependence and self-similarity network traffic. Authors experimented with Aucklands dataset. They used different time intervals of 1 ms and 10ms and 100 ms to test the model for prediction network traffic. Firstly, they applied ACF and PACF functions used to represent the states of time series. Then, ARIMA /GARCH model is applied to predict error network traffic at different time scales. A comparative prediction result between ARIMA/GARCH and FARIMA models is presented. They concluded non-linear time series ARIMA/GARCH model is better than linear time series FARIMA model.

Dehuai Zeng, Jianmin Xu et al. [21] proposed hybrid model of linear time series ARIMA and nonlinear time series Multilayer Artificial Neural Network (MLANN) to predict short term network traffic. They collected the data from Web server of Guangzhou network center after every 8 minute every day. Mean absolute relation and root means squared error (RMSE) functions are used to evaluate prediction performance of hybrid Model. A comparative prediction result between hybrid model and ARIMA and BPNN models is presented. They shared that the hybrid model achieved highest prediction than single mode.

S. Gowrishankar [18] studied the accuracy of linear time series nonlinear Fractional Auto Regressive Integrated Moving Average (FARIMA) and nonlinear time series model Recurrent Radial Basis Function Network (RRBFN) to forecast wireless network traffic for different time scales of 1, 10 second and 1 minute. Authors experimented with CRAWDAD dataset. A comparative between linear model and nonlinear model is presented. They shared that the non-linear RRBFN neural network techniques is slightly better to predict network traffic.

G.Rutka [65] used non-linear time series neural network techniques and linear time series ARIMA technique to predict self-similar network traffic. They evaluate their techniques, collected data set from free stats and Fotoblog websites for different time interval of two days. He applied Auto correlation function to represent time series with single step-head prediction and with multiple step-head prediction. A comparative prediction result between neural network and ARIMA approach is presented. He concluded that the ARIMA method is easy for training data set, but prediction result is not accurate and the neural network approach is quite complex but prediction result is very accurate.

Muhammad Faisal Iqbal et al. [30] combined linear time series ARIMA approach and non-linear time series (neural network) approach to predict network traffic.

They collected data set from Caida, Auckland, Bell core labs at different time intervals of varying size of data. They used NMSE metric to evaluate prediction performance of their techniques. A Comparison between linear techniques and non-linear techniques is presented. They noted that the non-linear neural network approach achieved better prediction of network traffic.

Chendi Feng, Hai-liong et al. [2] proposed non linear time series model of Linear Neural Network (LNN) and Elman neural network (ENN) approach to predict internet traffic. They applied wavelet method to decompose non stationary time series into several stationary components. Firstly, the simulation result for each model for network prediction is determined. Furthermore, the result of combination of LNN and ENN approaches with wavelet approaches is obtained. The normalized Means square error (NMSE) and single error ratio (SER) metrics are used to evaluate the prediction of their techniques. A comparative experimental result between individual model ENN and LNN and combined model (WLNN and WENN) is presented. They concluded that the hybrid model is more accurate for predicting network.

Mohamed Faten et al. [17] presented neural fuzzy model α- SNF, ARMA and ARIMA with time series models for enhancing quality of service of network traffic. They experimented on network data (Auckland, CESCA) obtained for different time intervals varying size of data. They divided the data set into two groups of training data set and other as testing data set. They proposed two steps to predict network first step, the training phase used to identify the model parameters. The testing of prediction used Root Means Square Error (RMSE) to measure prediction performance of α -SNF and ARIMA techniques. From the experimental result, they came with the choose input variable used Correlation Coefficient metrics to select suitable input variable desired with output variable. And also the suitable traffic granularity is varying from 100 ms to 200 Ms. Furthermore, the small training data less than 5 provide high prediction error. And Divide packets into small size give less prediction error.

LI Jing Fei, Shen Lei et al. [23] combined Wavelet and ARIMA models with time series for internet traffic analysis. They evaluated their technique, collected real data from the network of Shandong University for time intervals of five minute and three days. The Wavelet technique used to decompose the original sign into different frequency. They used approximation coefficient function to measure performance of Wavelet method. Further, they applied ARIMA model to predict error from network traffic. From the experiment result, they show that the combination of wavelet and ARIMA approach achieve more accuracy for prediction network traffic.

G. Rutka [9] described same aspect of linear time series ARMA and ARIMA and SARIMA approaches for modeling and forecasting of internet traffic. Author gathered network data set from free stats website. The free stats website includes data set at different time intervals varying size of data. He collected data set at different time interval 28 days and 7 days to detect the prediction of his approaches in long-rang dependence (LRD) and short rang dependence (SRD). Final Prediction Error (FPE) criterion metric is used to measure the prediction performance of his approaches. A Comparative prediction result between ARMA, ARIMA and SARIMA techniques are presented. They done that the SARIMA approach is better for prediction of internet traffic in SRD and LRD.

R.Sivakumar, E.ashok kumar et al. [56] proposed Hidden Markov Model (HMM) and neural network approaches with time series model to predict number of wireless devices connect to access point for enhancing quality of service. They experimented with data set from CRAWDAD (community resources archiving wireless data).They used transitive, probability metrics to evaluate the simulation result. A comparative prediction result between HMM model and neural approaches is presented. They shared that the neural network models is best suited for prediction network traffic.

Irina Klevecka [6] applied time series model with neural network and ARIMA approaches to predict short-rang network traffic. Authors investigate and evaluate their technique, collected data set from telephone network as well as packets switch IP network. They used root means square error (RMSE), means absolute error (MSE) and mean absolute percentage error (MAPE) metric to measure the performance of ARIMA approach for prediction. A comparative prediction result among ARIMA, neural network, Seasonal Exponential Smoothing and SARIMA approaches are presented. They noted that the neural network approach achieved better forecasting of short-rang network than another approaches.

### 3.4. *Decomposed model*

The time series generally decomposed into four components. Each component is defined as below.

3.4.1 Trend component

Trend is long –term propensity, increase and decrease in the time series data. Trend component represents the structural variations of low frequency time series.

3.4.2 Cyclical component

Cyclic pattern indicate the medium term fluctuation. The cyclical pattern display increase and falls without specified period.

3.4.3 Seasonal component

Seasonal component is variations in time series data that influenced by seasonal factors such as year, quarter, month, week, day, hour. The seasonal has stabled variation intra time series data.

3.4.4 Irregular component

Irregular component is residual time series after remove trend and seasonal. We display decomposed algorithms that are employed by researchers for network traffic predictions as follows.

Cheng Guang Jaian et al [3] proposed non-linear decomposed model by decomposed time series into trend component, period component, mutation component and random component for predicting long-rang network traffic. They evaluate their model, collected network data from backbone router of NSFNET with different period of time. The Auto correlation function used to measure simulation result. From experimental results, they compared decomposed model and ARIMA approach. They communicated that the decomposed model can get higher error precision to describe the long-rang traffic behavior.

### 3.5. *Evaluation metrics*

In network traffic prediction techniques many different metrics are used to investigate the quality of time series forecasting. The detection rate, false positive rate, accuracy and time cost metrics are employed for measuring the performance of classifier for different data set. A number of metrics exist obtained to express prediction accuracy. Each metric is defined as below.

(a) Mean Absolute Error (MAE )

It is the ratio of absolute value plus forecasting error divide by the number of forecasting.

(b) Mean Square Error (MSE)
It is the ratio of forecasting values plus actual values divided by the number of forecasting values.

(c) Root Mean Square Error (RMSE)
It is the ratio of observed values plus actual values divided by the number of forecasting values. The Root Mean Square Error (RMSE) (also called the root mean square deviation, RMSD) is a measure of the difference between values forecast by a model and the actual values observed from the environment. This means the RMSE is most helpful when large errors are mostly unfavorable.

(d) Normalized root mean square error (NRMSE)

It is the ratio of RMSE result divided by the of number maximum observed values plus minimum observed values of forecasting values.

(e) Mean percentage error (MPE)
It is the ratio of actual values plus forecasting value divided by forecasting values at a specific point of time series. MPE statistic is often changed by the mean absolute percentage error (MAPE).

(f) Mean Absolute Percentage Error (MAPE)
It is ratio of original values plus forecasting values divide by original values further divided by forecasting values. It is one of useful evaluation metrics that not influence by the numbers of the forecasting series.

Table 2. Network traffic prediction techniques

| Authors,years | Techniques | Data set | Purpose | Evaluation metrics |
|---|---|---|---|---|
| Nikkei C.anand et al. 2008 | GARCH | Abilene backbone internet | To predict and capture the bursty nature of Internet traffic | MLE |
| Edmund et al. 1993 | Neural network | Data set from B-ISDN. | Reduce noise video traffic in term of network traffic prediction | MSER |
| Dong –Churl et al.2009 | D-BLRNN | Bell core data set | To improve the BRNN approach for network traffic prediction | MSE |
| Samira Chabaa1 et al. 2010 | MLP | From backbone internet | To predict the internet traffic over IP network | RMSEP |
| Wang Junsong et al. 2009 | Elman Neural Network | Simulation | Long –rang internet traffic prediction | MSE,NMSE |
| Samira Chabaa et al. 2009 | ANFIS | From backbone internet over TCP/IP | For modeling and predicting internet traffic | RMSE, AARE |
| Paulo Cortez et al. 2006 | Neural Network Ensemble | From internet service provider on based TCP/IP protocol | To predict error from internet traffic | MAPE |
| Wang Peng et al. 2008 | BPWNN | Collected network dataset from real link | Improving drawback of BPNN technique in network traffic prediction | MSE |
| Hong Zhao 2009 | LMK | | To predict self-similar network traffic | LMS |
| S.M.Aqil et al. 2007 | ARIMA and GARCH | University of Calgary | To create conditional mean and conditional variance model for prediction internet traffic | R-squared, Adj.R-squared log likelihood f-statistic mean dependent var |
| BoZhou et al. 2006 | ARIMA/GARCH | From University of Auckland's Internet uplink in different | To predict busty nature of network traffic | SER |
| Dehuai Zeng et al. 2008 | ARIMA and MLANN | Web server of Guangzhou network centre | To predict error from network traffic | RMSE |
| G.Rutka 2008 | Neural network and ARIMA | Free stats and Fotoblog websites | To predict short-range and long-rang self-similar network traffic | FPE,AIC |
| Muhammad Faisal et al. 2012 | ARIMA and ARMA | From Caida, Auckland, Bell core data set | To predict error from network traffic | NMSE |
| Chendi Feng et al. | LNN and ENN | | To predict network traffic | NMSE,SER |
| Mohamed Faten et al. 2009 | (α-SNF), ARMA and ARIMA | Auckland, CESCA and Auckland | To predict network traffic | RMSE |
| LI Jing Fei et al. 2009 | Wavelet and ARIMA models | Shandong University | To predict network traffic | MAPE |
| R.Sivakumar et al. 2011 | HMM and NN | CRAWDAD dataset | To predict number of wireless devices connect to access point | Transitivity, probability |
| Irina Klevecka et al. 2009 | Neural network and ARIMA | Telephone network as well as packets switch IP network | To predict short-rang network traffic | RMSE,MSE,MAPE |
| Cheng Guang | Decomposed model | backbone router of NSFNET | To predict long-rang network traffic | AC |
| S. Gowrishankar | FARIMA and neural network | CRAWDAD website | To forecast wireless network traffic for improving quality of service (QoS) | MSE |
| N. K. Hong et al. 2011 | ARMA | Collected data set by using wire shack software | To predict FTP service in network traffic | MSE |
| Poo kuanHo et al. 2012 | ARMA | Collected 6 sets of real data at a period of six days from bit Torrent P2PN (point to point network by using wire shack software | To predict point-point network application such as bit torrent | MSE |
| Nayera Sadek et al. 2004 | k-Factor ARMA(autoregressive moving average) | Collected dataset from different real network data like MPEG, VIDEO, JPEG, INTERNT, and ETHERNT. | Predicting multi scale high speed network traffic by using different dataset . | MAE |
| Yanhua Yu et al. 2011 | APM | Certain mobile network of Heilongjian province in chain | To predict mobile network traffic | MAPE |
| Yanhua Yu et al. 2010 | ARIMA | Mobile network from Heilongjiang chain | To predict mobile network traffic | MAPE |
| Yantai Shu et al. 2003 | ARIMA | GSM wireless mobile real data from Tianjin at china mobile | To predict GMS wireless mobile network traffic | MMSE |
| Bozidar vujic et al. 2006 | SARIMA | Public safety trunked radio network | To predict call from radio network | NMS |
| Aqil Burney et al. 2009 | SARIMA | University of Karachi from High speed Fiber optics Local Area network with wireless computing | To reduce noise which happened in network traffic | MAE, MSE |
| Huda M.A.El Hag et al. 2007 | AARIMA | Start ware MPEG and Bell core | For predicting long-rang-dependant internet traffic | MAE |
| Shen Fu-ke et al. 2009 | EPTS | ECNU data set | Predict network traffic for managing bandwidth | FPE |

## 4. Conclusion

In the last decade, the analysis and prediction of network traffic has become a subject of continuous research in various sub-fields of computer networks. Innumerable number of researchers have been implemented an effective network traffic algorithm for the analysis and the prediction of network traffic.

In this paper, we surveyed the previous studies of network traffic analysis. We enlisted and discussed various approaches proposed to analyze and prediction of network traffic including data mining techniques, neural network and component analysis, and linear and nonlinear time series models. The tabular arrangement of all the surveyed papers shall give an overview of the research work in the analysis and prediction of network traffic.

Datasets used, implemented algorithms and metrics used to evaluate the results are also used to group the research works surveyed in the paper. Such a review paper would provide an insight about the topic to the new researchers.